\documentclass[sigconf]{acmart} 

\AtBeginDocument{
  }

\copyrightyear{2026}
\acmYear{2026}
\setcopyright{cc}
\setcctype{by}
\acmConference[ICSE '26]{2026 IEEE/ACM 48th International Conference on Software Engineering}{April 12--18, 2026}{Rio de Janeiro, Brazil}
\acmBooktitle{2026 IEEE/ACM 48th International Conference on Software Engineering (ICSE '26), April 12--18, 2026, Rio de Janeiro, Brazil}
\acmPrice{}
\acmDOI{10.1145/3744916.3773187}
\acmISBN{979-8-4007-2025-3/2026/04}

\usepackage{textcomp}
\usepackage{url}
\usepackage{algorithm}
\usepackage{algorithmic} 
\usepackage{listings}
\usepackage{multirow}
\usepackage{makecell}

\begin{document}

\title{TraceCoder: A Trace-Driven Multi-Agent Framework for Automated Debugging of LLM-Generated Code}

\author{Jiangping Huang}
\affiliation{%
  \institution{School of Computer Science and Technology, Chongqing University of Posts and Telecommunications}
  \city{Chongqing}
  \country{China}
}
\email{huangjp@cqupt.edu.cn}

\author{Wenguang Ye}
\affiliation{
  \institution{School of Computer Science and Technology, Chongqing University of Posts and Telecommunications}
  \city{Chongqing}
  \country{China}
}
\email{s231231083@stu.cqupt.edu.cn}

\author{Weisong Sun}
\authornote{Corresponding author.}
\affiliation{%
  \institution{Nanyang Technological University}
  \city{Singapore}
  \country{Singapore}
}
\email{weisong.sun@ntu.edu.sg}

\author{Jian Zhang}
\affiliation{%
  \institution{Beihang University}
  \city{Beijing}
  \country{China}
}
\email{zhangj3353@buaa.edu.cn}

\author{Mingyue Zhang}
\affiliation{%
  \institution{Southwest University}
  \city{Chongqing}
  \country{China}
}
\email{myzhangswu@swu.edu.cn}

\author{Yang Liu}
\affiliation{%
  \institution{Nanyang Technological University}
  \city{Singapore}
  \country{Singapore}
}
\email{yangliu@ntu.edu.sg}

\begin{abstract}
Large Language Models (LLMs) often generate code with subtle but critical bugs, especially for complex tasks. Existing automated repair methods typically rely on superficial pass/fail signals, offering limited visibility into program behavior and hindering precise error localization. In addition, without a way to learn from prior failures, repair processes often fall into repetitive and inefficient cycles. To overcome these challenges, we present TraceCoder, a collaborative multi-agent framework that emulates the observe-analyze-repair process of human experts. The framework first instruments the code with diagnostic probes to capture fine-grained runtime traces, enabling deep insight into its internal execution. It then conducts causal analysis on these traces to accurately identify the root cause of the failure. This process is further enhanced by a novel Historical Lesson Learning Mechanism (HLLM), which distills insights from prior failed repair attempts to inform subsequent correction strategies and prevent recurrence of similar mistakes. To ensure stable convergence, a Rollback Mechanism enforces that each repair iteration constitutes a strict improvement toward the correct solution. Comprehensive experiments across multiple benchmarks show that TraceCoder achieves up to a 34.43\% relative improvement in Pass@1 accuracy over existing advanced baselines. Ablation studies verify the significance of each system component, with the iterative repair process alone contributing a 65.61\% relative gain in accuracy. Furthermore, TraceCoder significantly outperforms leading iterative methods in terms of both accuracy and cost-efficiency.
\end{abstract}

\ccsdesc[500]{Software and its engineering~Software creation and management}
\ccsdesc[500]{Computing methodologies~Multi-agent systems}
\ccsdesc[300]{Software and its engineering~Software verification and validation}
\ccsdesc[300]{Computing methodologies~Natural language processing}

\keywords{Code Generation, Multi-Agent Systems, Self-Debugging, Runtime Tracing, Historical Lesson Learning, Large Language Models}

\maketitle

\section{Introduction}
\label{sec:introduction}

\begin{figure*}[ht]
  \centering
  \includegraphics[width=\textwidth]{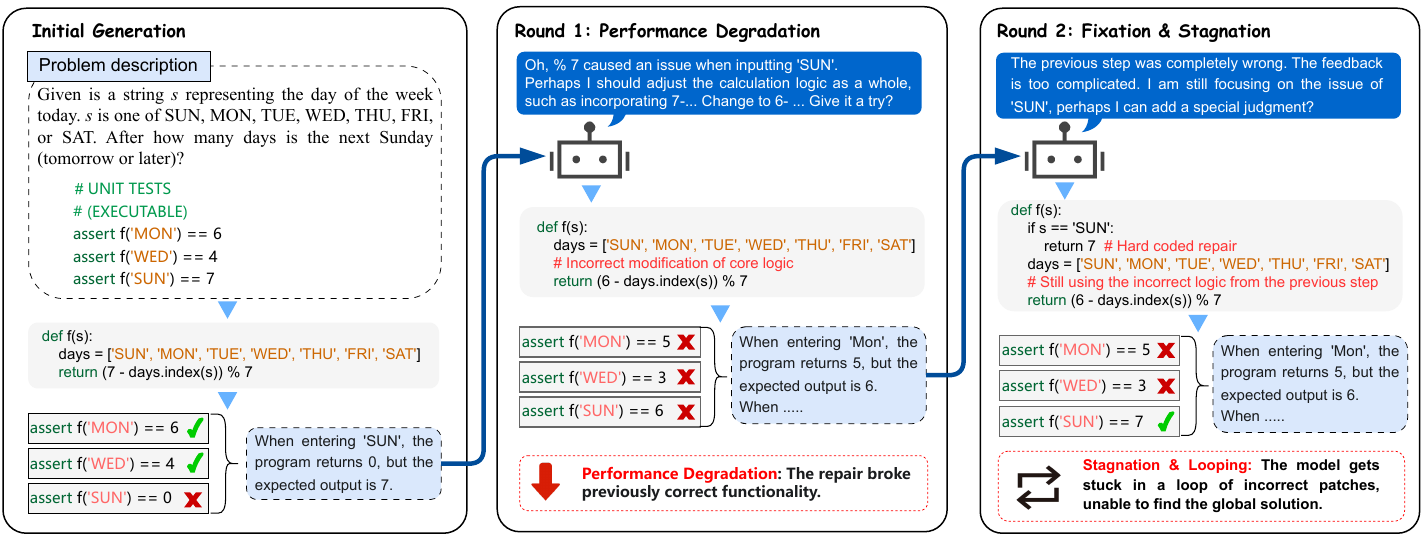}
  
  \Description[Limitations of simple execution feedback]{
    An illustration showing how relying solely on pass/fail execution feedback
    can mislead the model into producing increasingly incorrect patches.
    Without runtime insights, the model repeatedly makes local edits that
    worsen the program and fail to converge to a correct solution.
  }

  \caption{
    Limitations of simple execution feedback. Without runtime insights, the model repeatedly applies local patches that degrade the code's correctness, causing it to loop between incorrect versions rather than converging to a correct global solution.
  }

  \label{fig:limitations}
\end{figure*}

Large Language Models (LLMs)~\cite{chen2021evaluatinglargelanguagemodels, openai2023chatgpt, codellamaopenfoundation} have become increasingly powerful tools for software engineering tasks such as code generation~\cite{dong_self-collaboration_2024,mathews_test-driven_2024,han_archcode_2024}, code summarization~\cite{11029737,haldar_analyzing_2024,gu_assemble_2022}, and program  transformation~\cite{xian_transformcode_2024}. Despite their impressive capabilities, LLMs often generate code that contains subtle yet critical bugs—particularly in complex or logic-intensive scenarios~\cite{10850625,10992485}. This challenge has given rise to an emerging research direction focused on the automated repair of LLM-generated code, aiming to improve the reliability, correctness, and usability of LLM-assisted development~\cite{pornprasit_d-act_2023,dong_self-collaboration_2024}.

Recent work in this emerging area has explored diverse strategies for repairing LLM-generated code~\cite{10.1145/3631974}. These include using Chain-of-Thought (CoT)~\cite{NEURIPS2022_9d560961} prompting to guide repair reasoning~\cite{yin_thinkrepair_2024}, leveraging natural language as an intermediate representation for debugging~\cite{zhang-etal-2025-nl}, integrating statistical fault localization to improve fault awareness~\cite{fan_automated_2023,11029731}, proposing end-to-end multi-agent synergy for unified debugging~\cite{lee-etal-2025-unidebugger}, and fine-tuning LLMs for bug fixing~\cite{huang_empirical_2023}. Other approaches refine localization via token-level reasoning~\cite{hossain_deep_2024} or auxiliary fault-identification modules~\cite{krasniqi_hierarchical_2023}.

However, most existing self-correction methods operate as ``black-boxes'', relying solely on pass/fail feedback from a test suite. This approach, which lacks insight into the program's internal execution, suffers from significant limitations. As illustrated in Figure ~\ref{fig:limitations}, a simple execution-feedback model can easily fall into a degenerative cycle. In \textbf{Round 1}, deprived of detailed runtime information about \emph{why} a test failed, LLMs may make incorrect assumptions and apply faulty patches. This results in previously correct functionality being broken—referred to as \textbf{Performance Degradation}. Subsequently, in \textbf{Round 2}, the model may get stuck in a loop of incorrect local patches, leading to \textbf{Fixation \& Stagnation}, where it fails to diagnose the root cause and find the correct global solution. This example highlights two fundamental limitations in current LLM-based self-debugging methods: 1) they rely on binary final execution results, ignoring the rich semantics in intermediate execution states, which leads to imprecise fault localization; and 2) they adopt a stateless repair paradigm, unable to learn from historical debugging knowledge to avoid repeating past mistakes.

To address these challenges, we propose TraceCoder, a multi-agent collaborative automated debugging framework that emulates the human debugging process. Grounded in established cognitive models of expert debugging, which describe a cycle of \textit{information gathering}, \textit{hypothesis formation}, and \textit{repair}~\cite{mccauley2008debugging}, TraceCoder operationalizes this workflow through three specialized agents to enhance modularity, reliability, and control. Specifically, the Instrumentation Agent mirrors the information-gathering stage by capturing fine-grained runtime traces. The Analysis Agent then emulates hypothesis formation, performing causal reasoning on these traces, informed by a novel Historical Lesson Learning Mechanism (HLLM) that learns from past failures. The Repair Agent executes the resulting repair plan to modify the code. To ensure stable convergence, a Rollback Mechanism (RM) reverts the system to its last known correct state after any failed attempt. This structured, human-inspired workflow establishes a cohesive and interpretable debugging loop with a clear separation of concerns.

The proposed multi-agent collaborative automated debugging architecture confers several distinct advantages. First, it enables fine-grained runtime tracing, allowing the system to capture semantically rich execution signals. Second, it facilitates experience-informed repair decisions through historical learning, and ensures a robust repair trajectory via integrated rollback and iterative repair mechanisms. Furthermore, by decomposing the debugging task into logically interpretable stages, TraceCoder enhances modularity, explainability, and extensibility. We evaluate TraceCoder on three widely used datasets, including BigCodeBench~\cite{zhuo2024bigcodebench}, ClassEval~\cite{du_evaluating_2024}, and HumanEval+~\cite{liu_is_2023}, using diverse LLM backends. The results show that TraceCoder consistently outperforms existing automated debugging baselines, including Self-Debugging~\cite{chen2024teaching} and INTERVENOR~\cite{wang-etal-2024-intervenor}, measured by standard metrics such as Pass@1. Notably, TraceCoder improves the Pass@1 accuracy in repairing LLM-generated code, reduces redundant repair attempts, and enhances cost-efficiency, especially on complex programming tasks where LLMs are most prone to failure.

In summary, this paper makes the following contributions:

\begin{itemize}
    \item We present TraceCoder, a modular collaborative multi-agent framework that emulates the human debugging workflow to enable automated repair of LLM-generated code.
  
    \item We propose a novel HLLM that learns from prior failures to inform subsequent repairs and prevent recurring mistakes.
  
    \item Comprehensive evaluations demonstrate that TraceCoder substantially outperforms advanced baselines, achieving up to a 34.43\% relative improvement in Pass@1 accuracy on challenging class-level code generation benchmarks.
  
    \item We release an open-source implementation of TraceCoder~\cite{trace_coder} to support reproducibility and facilitate future research.
\end{itemize}

\section{Related Work}
\label{sec:related_work}

\subsection{Code Generation with LLMs}
\label{ssec:rw_llm_code_gen}
Recent advancements in LLMs have significantly propelled automated code generation. A growing body of work aims to improve the quality, correctness, and controllability of the generated code. For example, AMR-Evol~\cite{luo_amr-evol_2024} proposes a two-stage distillation framework to enhance generation fidelity, while ARCHCODE~\cite{han_archcode_2024} leverages in-context learning to translate software requirements into code and corresponding test cases. Empirical investigations, such as DevGPT~\cite{jin_can_2024}, reveal that LLM-generated code is often used for prototyping or conceptual illustration, rather than deployment. To support more rigorous evaluation, several benchmarks have been introduced: DA-Code~\cite{huang_da-code_2024} focuses on agent-based workflows, ClassEval~\cite{du_evaluating_2024} targets class-level programs with structural dependencies, and EvalPlus~\cite{liu_is_2023} augments test suites to thoroughly assess functional correctness. On the algorithmic front, PG-TD~\cite{zhang_planning_2023} incorporates planning-guided decoding with lookahead search, while a self-planning framework~\cite{10.1145/3672456} decomposes intent into subgoals to improve generation reliability. Despite these advances, most work focuses solely on improving generation itself, leaving post-generation debugging comparatively underexplored. LLM-generated code still frequently contains subtle logic errors that existing generation pipelines cannot reliably detect or correct. This gap motivates our work: we introduce a trace-driven, multi-agent framework that leverages runtime evidence and iterative refinement to not only diagnose but also systematically repair LLM-generated code.

\subsection{Automated Program Repair}
\label{ssec:rw_apr}
Automated Program Repair (APR) is a long-standing field in software engineering focused on automatically fixing bugs in source code. Traditional APR techniques often rely on search-based methods~\cite{le2011genprog}, which use genetic algorithms to evolve patches, or template-based approaches~\cite{kim2013automatic}, which apply predefined fix patterns. These methods have been extensively evaluated on benchmarks like Defects4J~\cite{just2014defects4j}, which contains real-world bugs from large-scale Java projects. However, they often struggle with complex logical errors that require a deep semantic reasoning about program behavior.

The advent of LLMs has opened new frontiers for APR. Recent works such as RepairAgent~\cite{Islem_icse_2025} and ThinkRepair~\cite{yin_thinkrepair_2024} leverage LLMs to reason about bugs and synthesize human-like patches, often outperforming traditional methods. These approaches typically adopt an iterative feedback loop, where the model refines the code based on test outcomes. For instance, Self-Debugging~\cite{chen2024teaching} uses simple pass/fail signals, while INTERVENOR~\cite{wang-etal-2024-intervenor} employs a dual-agent "teacher-learner" framework to guide the repair process. A closely related work, AutoVerus~\cite{yang_AutoVerus_2025}, also uses a feedback loop for repairing Rust programs but focuses specifically on resolving formal verification errors, where feedback is structured and precise.

TraceCoder extends this line of work with three key distinctions. First, while most methods rely on black-box test outcomes, TraceCoder leverages fine-grained runtime traces, providing white-box visibility into program behavior for more precise fault localization. Second, unlike AutoVerus, which targets formal verification, TraceCoder is designed for general-purpose code and unstructured test feedback, a common and challenging scenario in practice. Most importantly, our novel HLLM addresses an overlooked gap by enabling the system to learn from past failures, preventing repeated mistakes and improving efficiency in repairing complex bugs.

\subsection{LLM-Based Multi-Agent Systems}
\label{ssec:rw_multi_agent_collab}
While monolithic LLMs have achieved strong performance across a variety of tasks, they often struggle in scenarios requiring complex strategic reasoning, iterative refinement, and dynamic adaptation. Limitations such as fixed context windows and unidirectional generation hinder their ability to perform trial-and-error reasoning, reflect on prior work, or plan over long horizons~\cite{NEURIPS2022_9d560961, chen2021evaluatinglargelanguagemodels}. To overcome these challenges, recent research has turned toward collaborative multi-agent systems (MAS), where multiple role-specialized agents interact to decompose complex problems and simulate human collaborative behavior~\cite{wang2024survey,huang2025envisioning}. Systems such as AgentVerse~\cite{ICLR2024_578e65cd} and OpenDevin~\cite{opendevin2024} demonstrate how agent-based collaboration, when supported by robust communication protocols and task-aligned workflows, can lead to emergent capabilities, achieving results that surpass the performance of individual agents~\cite{riedl2025emergentcoordinationmultiagentlanguage}. Advances in multi-agent alignment~\cite{lin-etal-2025-creativity}, reflective reasoning mechanisms~\cite{bo2024reflective}, and agent-oriented planning strategies~\cite{li2025agentoriented} further enhance reliability across multiple rounds of iterative decision-making. However, most MAS frameworks focus on task decomposition and static role assignment, with limited support for integrating dynamic runtime feedback or leveraging accumulated debugging history, both of which are critical for effective automated program repair. We extend this line of work by introducing a multi-agent architecture that fuses causal planning with collaborative repair to support runtime-aware, history-informed, and self-correcting debugging.

\begin{figure*}
  \centering
  \includegraphics[width=\textwidth]{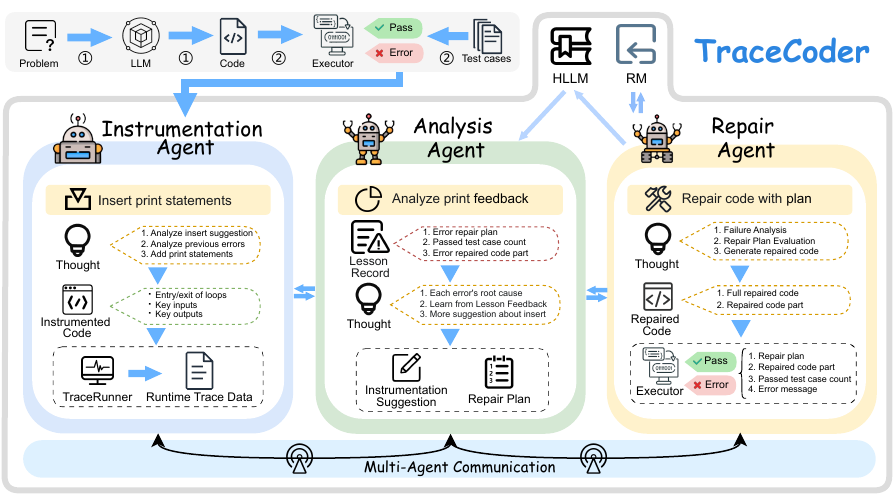}
  \Description[TraceCoder overall workflow diagram]{
    Workflow for TraceCoder: (1) LLM generates initial code; (2) code is executed and tested;
    then a multi-agent loop (Instrumentation Agent, Analysis Agent, Repair Agent) uses runtime tracing,
    high‑level LLM and repair model to simulate expert debugging and iteratively improve code.
    The HLLM records failed attempts from the Repair Agent and provides lessons to the Analysis Agent in the next iteration.
  }
\caption{
Overview of TraceCoder’s workflow. 
\textcircled{1} An LLM generates an initial code solution. 
\textcircled{2} The code is executed and tested. 
A multi-agent debugging loop—comprising the Instrumentation, Analysis, and Repair Agents—emulates expert debugging behaviors by leveraging runtime tracing, HLLM, and RM to enable effective and stable repair. 
After each failed attempt, the HLLM logs the outcome and informs the Analysis Agent’s strategy for the subsequent cycle.
}
\label{fig:tracelearncoder_workflow}
\end{figure*}

\section{Methodology}
\label{sec:methodology}

\subsection{Overview}
\label{ssec:overview}
We introduce TraceCoder, a trace-driven automated debugging framework that iteratively repairs LLM-generated code by emulating expert debugging workflows, as illustrated in Figure~\ref{fig:tracelearncoder_workflow}. The framework is organized around three specialized agents: the Instrumentation Agent inserts diagnostic probes to collect fine-grained runtime traces; the Analysis Agent performs causal reasoning over these traces to localize faults; and the Repair Agent synthesizes and applies concrete code modifications. To guide this process, TraceCoder integrates the HLLM to learn from past repair failures, and the RM to ensure stable convergence. When initial code fails its test suite, these agents are activated and iterate until all tests pass or a termination condition is reached.

\subsection{Instrumentation Agent}
\label{ssec:instrumentation_agent}
The Instrumentation Agent is a core component of the proposed framework, responsible for collecting dynamic execution information. Its internal \textit{thought process} involves lightweight reasoning over previous execution failures and current instrumentation suggestions to determine suitable probe locations. Based on this reasoning, the agent inserts diagnostic print statements into the code to reveal internal state transitions and control-flow behavior. These runtime insights serve as essential evidence for downstream causal analysis by the Analysis Agent. The Instrumentation Agent is invoked in two scenarios: when the initial execution fails its test suite, or when a subsequent repair attempt does not resolve the issue.

Its inputs are context-dependent, including: (1) the current code (\(C_{\text{faulty}}\)) to be instrumented, (2) the most recent test failure feedback (\(F_{\text{error}}\)) from a failed repair attempt, and (3) optionally, a set of fine-grained instrumentation suggestions (\(I_{\text{sugg}}\)) from the Analysis Agent. By synthesizing this contextual information, the Instrumentation Agent generates a new version of the code augmented with diagnostic probes. The resulting instrumented code (\(C_{\text{inst}}\)) faithfully preserves the original computational semantics but emits context-aware debug logs during execution, providing valuable insights into the program’s dynamic behavior. The core operation of the Instrumentation Agent can be summarized as:
\begin{equation}
(C_{\text{faulty}}, F_{\text{error}},  I_{\text{sugg}}) \rightarrow C_{\text{inst}}
\label{eq:instrumentation_agent}
\end{equation}

The Instrumentation Agent employs a dedicated prompt that directs the LLM to strategically insert diagnostic probes into the faulty code. 
The prompt instructs the model to augment Python functions with probes guided by the test failure feedback, 
revealing both execution flow and key variable states. 
It explicitly encourages failure-aware instrumentation by prioritizing regions relevant to the observed failure, 
thereby avoiding indiscriminate logging and enhancing both the efficiency and informational value of runtime traces. 
To ensure consistency and effectiveness, the agent adheres to four core principles:
\begin{itemize}
    \item \textbf{Logical Decomposition.} The code is decomposed into distinct logical units, such as function bodies, branches, and loops, that define the instrumentation scope.
    \item \textbf{State and Control Traceability.} Each unit is augmented with print statements that log key inputs, outputs, and intermediate values. Entry and exit points of major blocks are also traced to reveal control flow.
    \item \textbf{Instrumentation Purity.} Only non-invasive print statements may be inserted. The agent must not modify computational logic, comment out code, or introduce new variables, thus preserving semantic integrity.
    \item \textbf{Readable and Structured Output.} All logs must follow a clean and structured format to facilitate downstream analysis by both humans and automated tools.
\end{itemize}

Ensuring semantic preservation is critical for reliable instrumentation. Although formal guarantees of LLM behavior are inherently difficult, we enforce this constraint through strictly designed prompts that prohibit any modifications to the program’s logic, control flow, or data structures, allowing only the insertion of `print` statements. To assess the effectiveness of this enforcement, we conducted an empirical validation study (Section~\ref{sssec:validation_purity}), which demonstrates a semantic preservation rate of over 99\%.

After generating the instrumented code, the Instrumentation Agent proceeds to the dynamic execution phase by submitting the code and its associated test suite to \textit{TraceRunner}, a custom execution wrapper. TraceRunner performs controlled execution by isolating runtime environments and capturing both diagnostic logs and exception traces in a structured format. The resulting runtime trace is forwarded to the Analysis Agent as a key diagnostic artifact, enabling it to perform causal reasoning and plan targeted repair actions for the next iteration.

\subsection{Analysis Agent}
\label{ssec:analysis_agent}

The Analysis Agent diagnoses program failures by integrating runtime traces and historical failure data. It performs context-aware reasoning to identify root causes and derive actionable debugging insights across iterations, producing two complementary outputs: a repair plan for the Repair Agent and targeted instrumentation suggestions for the next debugging cycle. Formally, it operates on a structured set of inputs:

\begin{itemize}
    \item \textbf{Original Problem Description} (\(D_{\text{prob}}\)). 
    A textual summary of the program’s intended functionality and expected behavior, serving as a semantic reference for diagnosis.
    \item \textbf{Instrumented Code} (\(C_{\text{inst}}\)). The latest version of the source code augmented with diagnostic probes, generated by the Instrumentation Agent for runtime behavior analysis.
    \item \textbf{Runtime Trace Data} (\(T_{\text{runtime}}\)). Fine-grained execution logs captured by \textit{TraceRunner}, including diagnostic outputs and all recorded runtime errors or exceptions.
    \item \textbf{Lesson Record} (\(L_{\text{record}}\)). A structured log of all failed repair attempts for the current problem, used to reflect on prior reasoning and avoid repeated mistakes.
\end{itemize}

The diagnostic capability of the Analysis Agent is driven by a structured \textit{thought process}, implemented through a prompt-based reasoning schema. This prompt guides the agent to: (1) analyze runtime traces to identify the root cause of the error; (2) incorporate historical insights via the HLLM to avoid repeated mistakes; and (3) generate targeted instrumentation suggestions to guide future data collection. This diagnostic workflow is implemented as a two-stage procedure:

\begin{itemize}
    \item \textbf{Diagnosis and Reflection.} The agent analyzes the Runtime Trace Data to identify the root cause of failure, while concurrently analyzing the Lesson Record to understand why previous attempts were unsuccessful. This reflective process enables the agent to learn from previously flawed reasoning steps, not just surface-level code issues.
    \item \textbf{Strategy Formulation.} Based on its diagnosis, the agent produces two complementary outputs:
    \textbf{(1)} \textit{Repair Plan} (\(P_{\text{repair}}\)). A proposed code modification plan passed to the Repair Agent for implementation.
    \textbf{(2)} \textit{Instrumentation Suggestion} (\(I_{\text{sugg}}\)). Targeted guidance for the next debugging cycle, specifying where new probes should be inserted to verify the repair or collect additional runtime signals if it fails.
\end{itemize}

This procedure is formalized as a function that maps diagnostic inputs to actionable repair and instrumentation outputs:

\begin{equation}
(D_{\text{prob}}, C_{\text{inst}}, T_{\text{runtime}}, L_{\text{record}}) \rightarrow (P_{\text{repair}}, I_{\text{sugg}})
\label{eq:analysis_agent}
\end{equation}

\subsection{Repair Agent}
\label{ssec:repair_agent}

The Repair Agent serves as the primary executor in the TraceCoder framework, responsible for translating high-level repair plans into concrete code modifications. Its core function is to implement the changes proposed by the Analysis Agent and participate in the feedback loop by reporting failure cases that guide future refinement. To support this task, the Repair Agent operates on four key inputs:
\begin{itemize}
    \item \textbf{Original Problem Description.} A complete textual specification of the programming task, providing semantic and functional context for the intended behavior.
    \item \textbf{Code to Be Repaired.} The latest non-instrumented version of the source code from the current iteration that failed testing and still requires correction.
    \item \textbf{Test Failure Feedback.} Diagnostic feedback summarizing test case failures, including observed outputs, error messages, and assertion violations during execution.
    \item \textbf{Structured Repair Plan.} A detailed, step-by-step modification strategy generated by the Analysis Agent, explicitly specifying how the faulty code should be corrected.
\end{itemize}

The core mechanism of the Repair Agent centers on its interaction with LLMs, guided by a carefully designed prompt that embodies its internal \textit{thought process}. This prompt defines the LLM's role, clarifies the repair objectives, and guides it to reason through the task in a systematic and controlled manner. To fulfill its function, the Repair Agent follows a structured three-step workflow:
\begin{itemize}
    \item \textbf{Failure Analysis.} 
    The LLM first performs a detailed analysis of the provided test failure feedback to precisely identify and diagnose the root cause of the error.
    \item \textbf{Repair Plan Evaluation.} It then rigorously evaluates the repair plan proposed by the Analysis Agent to assess whether the suggested modifications are both logically sound and sufficient to address the identified issue.
    \item \textbf{Code Repair Execution.} Finally, the LLM applies the validated repair plan to modify the code. This step requires strict adherence to the specified changes to ensure the defect is correctly resolved without introducing new errors.
\end{itemize}

To enhance the robustness and success rate of the repair process, the instructional prompt provides the LLM with a controlled degree of flexibility. If, after evaluating the repair plan, the LLM identifies minor omissions that do not compromise the core strategy, it is allowed to make localized, minor code modifications (e.g., correcting a variable name or adjusting a boundary condition) that do not contradict the core strategy of the Analysis Agent's repair plan. This ensures the issue can be effectively resolved while maintaining high fidelity to the Analysis Agent’s intent. Formally, the operation of the Repair Agent can be modeled as:
\begin{equation}
(D_{\text{prob}}, C_{\text{faulty}}, F_{\text{error}}, P_{\text{repair}}) \rightarrow C_{\text{repaired}}
\label{eq:repair_agent}
\end{equation}

\subsection{Multi-Agent Communication}
\label{ssec:multi_agent_communication}

The communication among TraceCoder's agents follows a structured, sequential pattern, mediated by shared artifacts rather than direct message passing. This design supports a disciplined, cyclical repair process. When a repair cycle begins, the Instrumentation Agent receives the faulty code along with its corresponding test failure feedback. It produces an instrumented version of the code, which is executed in the \textit{TraceRunner} environment. The resulting runtime traces serve as the primary input to the Analysis Agent.

The Analysis Agent integrates these traces with the original problem description and historical repair experiences to diagnose the root cause of failure. It generates two outputs: a detailed repair plan for the Repair Agent, and instrumentation suggestions to guide the Instrumentation Agent in subsequent iterations. This feedback loop enables progressively refined analysis. Upon receiving the repair plan, the Repair Agent applies the specified modifications, and the new candidate solution is re-tested, initiating the next cycle.

This sequential, artifact-mediated communication model ensures that each agent operates with well-structured, contextually relevant information. We opt for this custom control loop over a dedicated multi-agent system (MAS) framework (e.g., AgentVerse) because our workflow is linear and deterministic. For such a structured process, a full-fledged MAS framework would introduce unnecessary complexity and overhead without providing significant benefits. Our direct approach maintains full control and enhances reproducibility, which is crucial for scientific evaluation.

Importantly, after each failed repair attempt, the results are logged into the HLLM, which distills summarized lessons and provides them back to the Analysis Agent in the subsequent iteration. This mechanism closes the learning loop and enables history-informed, progressively refined debugging.

\subsection{Historical Lesson Learning Mechanism}
\label{ssec:failure_experience_mechanism}

A key component of TraceCoder is the HLLM, which addresses the limitations of stateless repair by enabling the system to learn from past failures. Inspired by the Trial-and-Error Learning Theory~\cite{thorndike1913educational}, HLLM systematically records, retrieves, and reuses failed repair attempts from prior iterations, referred to as \textit{Lesson Feedback}, on the same problem instance. This allows the Analysis Agent to avoid previously ineffective reasoning paths and refine its diagnostic approach across repair cycles. By leveraging these historical insights, HLLM improves debugging efficiency, reduces redundant attempts, and increases correction success rates in complex scenarios. The mechanism operates through the following three stages.

\subsubsection{Lesson Record}
Each time an iterative repair attempt fails to pass all predefined test cases, the system automatically captures key contextual information. This is recorded as an execution result (\(E_{\text{result}}\)) and a detailed execution message (\(E_{\text{message}}\)). If the execution fails, this message contains the specific repair plan that was attempted (\(P_{\text{repair}}\)), the resulting faulty code (\(C_{\text{repaired}}\)), the error feedback from the failed execution (\(F_{\text{error}}\)), and the passed test case count (\(S_{\text{passed}}\)). These records collectively constitute the Lesson Feedback for the current specific problem.

\subsubsection{Lesson Feedback}
Before generating a new repair plan, the Analysis Agent prompts the LLM to analyze the Lesson Record, which aggregates all failure records for the current problem instance. From this analysis, the LLM obtains a Lesson Feedback, allowing it to avoid previously ineffective strategies and make more informed repair decisions based on past diagnostic experiences.

\subsubsection{Lesson-Informed Deliberation and Planning}
The Analysis Agent is explicitly guided, via its structured prompt, to conduct deliberate reasoning over the retrieved Lesson Feedback before generating a new repair plan. This process involves three key tasks:
\begin{itemize}
    \item Diagnosing the root causes of previous repair failures to understand why earlier strategies were ineffective;
    \item Summarizing recurring pitfalls or suboptimal repair patterns identified across multiple failed attempts;
    \item Formulating a revised repair plan that aims to address prior deficiencies and to explore alternative solutions.
\end{itemize}

To operationalize the synergy across the three stages, Algorithm~\ref{alg:hllm} illustrates the core workflow of the HLLM. As shown, the algorithm processes a structured $E_{\text{message}}$ object, which encapsulates the full context of a failed repair attempt. It then extracts failure patterns from this context to synthesize structured lessons, which in turn guide the formulation of subsequent repair strategies.

\begin{algorithm}
\caption{Historical Lesson Learning Mechanism}
\label{alg:hllm}
\begin{algorithmic}[1] 
    \REQUIRE Result of code execution, $E_{\text{result}}$ \\
             Message of code execution, $E_{\text{message}}$ \\
             The historical lesson record, $L_{\text{record\_in}}$ 
    \ENSURE The updated lesson record, $L_{\text{record\_out}}$ 

    \STATE $L_{\text{record\_updated}} \leftarrow L_{\text{record\_in}}$ 
  
    \IF{$E_{\text{result}}$ is a failure}
        \STATE Extract $P_{\text{repair}}$ from $E_{\text{message}}$
        \STATE Extract $F_{\text{error}}$ from $E_{\text{message}}$
        \STATE Extract $C_{\text{repaired}}$ from $E_{\text{message}}$
        \STATE Extract $S_{\text{passed}}$ from $E_{\text{message}}$
      
        \STATE new\_record $\leftarrow$ ($P_{\text{repair}}$, $F_{\text{error}}$, $C_{\text{repaired}}$, $S_{\text{passed}}$) 
      
        \STATE \textbf{Add} new\_record to $L_{\text{record\_updated}}$ 
    \ENDIF

    \STATE $L_{\text{record\_out}} \leftarrow L_{\text{record\_updated}}$ 
    \RETURN $L_{\text{record\_out}}$
\end{algorithmic}
\end{algorithm}

\subsection{Rollback Mechanism}
\label{ssec:rollback_mechanism}

To maintain both convergence and robustness throughout the iterative self-debugging process, TraceCoder incorporates the RM. This mechanism serves as a critical strategy for state management and recovery, designed to revert the system to a previously validated or superior state whenever a new repair attempt fails to yield progress or introduces regressions. By doing so, RM prevents the repair trajectory from deteriorating across iterations and anchors the search process around the best-known solutions. The operation of the RM is structured around the following two core procedures.
\subsubsection{Key State Recording}
Throughout the debugging process, TraceCoder continuously tracks several key state variables to inform its decisions. It maintains a record of the historically best-performing code so far (\(C_{\text{best}}\)), defined as the version that has passed the greatest number of test cases. Its corresponding score, the highest passed test count so far (\(S_{\text{best}}\)), is stored as the primary performance benchmark. To detect significant regressions, TraceCoder also records the passed test count of the previous attempt (\(S_{\text{previous}}\)). Finally, a stagnation counter (\(k\)) is maintained to track the number of consecutive attempts that have failed to improve the best score.
\subsubsection{Progress Evaluation and Decision-Making}
Each time the Repair Agent generates a new repair candidate  ($C_{\text{attempted}}$), TraceCoder evaluates its performance by obtaining its passed test count ($S_{\text{attempted}}$). This score is then compared against the historically best score ($S_{\text{best}}$) and the previous attempt's score ($S_{\text{previous}}$) to make a decision, as formalized in Algorithm~\ref{alg:rollback_decision}. The algorithm returns an updated state tuple $\langle C_{\text{best\_new}}, S_{\text{best\_new}}, C_{\text{next\_base}}, k_{\text{new}} \rangle$, where $C_{\text{next\_base}}$ specifies the baseline code (either $C_{\text{attempted}}$ or $C_{\text{best}}$) for the subsequent repair cycle. The decision process operates through three distinct scenarios.
\begin{itemize}
    \item \textbf{Improvement.} If the new code passes more test cases than the current best, it is promoted as the new historically best, and the next iteration proceeds from it. The counter for non-improving attempts is reset.
    \item \textbf{Stagnation or Regression.} If the new code shows no improvement or performs worse, the system reverts to the previously recorded best version for the next attempt.
    \item \textbf{Prolonged Stagnation.} If no progress occurs after several iterations, the repair process is terminated to prevent wasted computation and potential overfitting.
\end{itemize}

Building on the two core processes above, the decision logic of the RM is formalized in Algorithm~\ref{alg:rollback_decision}. It performs precise state management by quantitatively comparing the performance of the new candidate against the historically best version, while monitoring for repeated setbacks. This ensures that the repair process consistently progresses toward an optimal solution.

\begin{algorithm}
\caption{Rollback Mechanism}
\label{alg:rollback_decision}
\begin{algorithmic}[1]
    \REQUIRE The latest attempted code, $C_{\text{attempted}}$ \\
             The best-performing code so far, $C_{\text{best}}$ \\
             The highest passed test count so far, $S_{\text{best}}$ \\
             The passed test count of the previous attempt, $S_{\text{previous}}$ \\
             The stagnation counter, $k$
    \ENSURE The decision: $\langle \text{accept}, \text{continue}, \text{rollback} \rangle$ \\
             The updated state: $\langle C_{\text{best\_new}}, S_{\text{best\_new}}, C_{\text{next\_base}}, k_{\text{new}} \rangle$
    \STATE Let $S_{\text{attempted}}$ be the number of tests passed by $C_{\text{attempted}}$
    \STATE $\Delta \leftarrow S_{\text{attempted}} - S_{\text{best}}$
    
    \IF{$\Delta > 0$}
        \STATE $C_{\text{next\_base}} \leftarrow C_{\text{attempted}}$ \hfill // Promote new code
        \RETURN $\langle \text{accept}, \langle C_{\text{attempted}}, S_{\text{attempted}}, C_{\text{next\_base}}, 0 \rangle \rangle$
    \ELSIF{$\Delta = 0$}
        \STATE $C_{\text{next\_base}} \leftarrow C_{\text{attempted}}$ \hfill // Continue with current
        \RETURN $\langle \text{continue}, \langle C_{\text{best}}, S_{\text{best}}, C_{\text{next\_base}}, k+1 \rangle \rangle$
    \ELSE
        \IF{$S_{\text{attempted}} < S_{\text{previous}}$}
            \STATE $C_{\text{next\_base}} \leftarrow C_{\text{best}}$ \hfill // Trigger Rollback
            \RETURN $\langle \text{rollback}, \langle C_{\text{best}}, S_{\text{best}}, C_{\text{next\_base}}, k+1 \rangle \rangle$
        \ELSE 
            \STATE $C_{\text{next\_base}} \leftarrow C_{\text{attempted}}$ \hfill // Stagnation, keep trying
            \RETURN $\langle \text{continue}, \langle C_{\text{best}}, S_{\text{best}}, C_{\text{next\_base}}, k+1 \rangle \rangle$
        \ENDIF
    \ENDIF
\end{algorithmic}
\end{algorithm}

\section{Evaluation}
\label{sec:evaluation}

This section presents four research questions addressed by TraceCoder and details the experimental setup, including datasets, baselines, evaluation metrics, and implementation details. We conduct extensive experiments to answer these questions and provide a comprehensive evaluation of TraceCoder’s effectiveness.

\subsection{Research Questions}
\label{ssec:research_questions}
\begin{enumerate}
    \item[\textbf{RQ1:}] How effective is TraceCoder at repairing LLM-generated code compared to advanced automated repair methods?
  
    \item[\textbf{RQ2:}] How do TraceCoder's key hyperparameters affect its repair performance and stability?
  
    \item[\textbf{RQ3:}] What is the contribution of each core component to TraceCoder’s overall effectiveness?
  
    \item[\textbf{RQ4:}] How does TraceCoder perform in practice, particularly compared to sampling-based strategies, in terms of reliability, cost efficiency, and failure modes?
\end{enumerate}

\subsection{Experimental Setup}
\label{ssec:experimental_setup}

\subsubsection{Datasets}
\label{sssec:datasets}

We conduct a comprehensive evaluation of TraceCoder on four widely adopted benchmark datasets: \textit{HumanEval}~\cite{chen2021evaluatinglargelanguagemodels}, \textit{HumanEval+}~\cite{liu_is_2023}, \textit{BigCodeBench}~\cite{zhuo2024bigcodebench}, and \textit{ClassEval}~\cite{du_evaluating_2024}. The tasks in \textit{HumanEval}, \textit{HumanEval+}, and \textit{BigCodeBench} are at the function level, where the goal is to generate a single correct Python function. \textit{HumanEval+} extends \textit{HumanEval} with broader and more robust test coverage. \textit{BigCodeBench} offers a diverse set of realistic function-level tasks emphasizing complex instruction following. By contrast, \textit{ClassEval} targets class-level code generation, evaluating LLMs on object-oriented constructs such as inter-method dependencies and class hierarchies. To mitigate test leakage, we ensured that \textit{BigCodeBench} was released after the knowledge cut-off date of the evaluated LLMs, avoiding prior exposure during training.

\subsubsection{Baselines}
\label{sssec:baselines}
To validate its effectiveness, we compare TraceCoder with five representative baseline methods that reflect different paradigms in LLM-based code generation and repair.
\begin{itemize}
    \item \textbf{Direct}~\cite{chen2021evaluatinglargelanguagemodels}: Generates code directly from the problem description without additional reasoning or planning.
  
    \item \textbf{CoT}~\cite{NEURIPS2022_9d560961}: Encourages the LLM to produce intermediate reasoning steps before code generation.

    \item \textbf{Self-Planning}~\cite{10.1145/3672456}: Decomposes complex problems into a series of subgoals, improving task structure and clarity.
  
    \item \textbf{Self-Debugging}~\cite{chen2024teaching}: Executes the generated code and uses its execution results to iteratively refine the solution.
  
    \item \textbf{INTERVENOR~\cite{wang-etal-2024-intervenor}}: Improves repair accuracy by alternating LLM roles as learner and teacher, simulating human-like interactions to produce a Chain-of-Repair.
\end{itemize}

\subsubsection{Metrics}
\label{sssec:metrics}
We evaluate functional correctness using the standard \textit{Pass@K} metric~\cite{10.1145/3597503.3623316}, which measures the percentage of problems for which at least one of $k$ generated solutions passes all benchmark test cases. In our experiments, we adopt the greedy \textbf{Pass@1} setting ($k=1$), where only a single solution is evaluated per problem. This stricter metric better reflects real-world development scenarios, as developers typically do not sample multiple candidates but rely on the top-1 generated solution.

\subsubsection{Implementation details}
\label{sssec:models_used}
We evaluate TraceCoder using three representative LLMs: \textit{Gemini-2.5-flash-0417}~\cite{geminiteam2024gemini15unlockingmultimodal}, \textit{DeepSeek-V3-0324}~\cite{deepseekai2025deepseekv3technicalreport}, and \textit{Qwen-Plus-2025-01-25}~\cite{qwen2025qwen25technicalreport}. For any given run, a single LLM is used consistently across all agents and the initial code generation module. To ensure result reproducibility, we adopt default deterministic API configurations when invoking the models.

For a fair comparison, all baseline methods are implemented using the same base LLMs and API parameters as TraceCoder. We reproduce each baseline based on its official implementation or prompt design. During initial code generation, all methods, including TraceCoder, generate code based solely on the natural language problem description; the test suite is used exclusively for verification and feedback in subsequent repair stages. We limit all iterative methods, including TraceCoder and the iterative baselines, to a maximum of 5 repair attempts. All generated code is executed and evaluated in a unified Python 3.10 environment.

\begin{table*}
\centering
\caption{Comparison of Pass@1 (\%) between TraceCoder and baseline methods across four benchmarks and three foundation models. ``Ours'' denotes the proposed TraceCoder. Bold numbers indicate the best performance in each column; values in parentheses denote the relative improvement ($\uparrow$\%) over the second-best result.}
\label{tab:main_results_split}
\resizebox{\textwidth}{!}{%
\begin{tabular}{@{}llcccccc@{}}
\toprule
\textbf{Models} & \textbf{Methods} & \textbf{HumanEval} & \textbf{HumanEval+} & \textbf{ClassEval} & \textbf{BigCodeBench-Complete} & \textbf{BigCodeBench-Instruct} & \textbf{Average} \\ \midrule
\multirow{6}{*}{\begin{tabular}[c]{@{}l@{}}Gemini-2.5-Flash\\-0417\end{tabular}} & Direct & 96.34 & 91.46 & 38.00 & 53.77 & 43.77 & 64.67 \\
 & CoT & 93.90 & 91.46 & 41.00 & 53.86 & 43.68 & 64.78 \\
 & Self-Planning & 94.51 & 90.85 & 36.00 & 55.61 & 43.15 & 64.02 \\
 & Self-Debugging & 98.78 & 96.34 & 61.00 & 78.07 & 71.05 & 81.05 \\
 & INTERVENOR & \textbf{99.39} & 95.12 & 61.00 & 75.88 & 69.82 & 80.24 \\
 & \textbf{Ours} & \textbf{99.39} & \textbf{98.17 ($\uparrow$\,1.90\%)} & \textbf{82.00 ($\uparrow$\,34.43\%)} & \textbf{89.04 ($\uparrow$\,14.05\%)} & \textbf{85.00 ($\uparrow$\,19.63\%)} & \textbf{90.72 ($\uparrow$\,11.93\%)} \\ \midrule
\multirow{6}{*}{\begin{tabular}[c]{@{}l@{}}DeepSeek-V3\\-0324\end{tabular}} & Direct & 94.51 & 90.24 & 41.00 & 38.25 & 46.67 & 62.13 \\
 & CoT & 93.29 & 88.41 & 41.00 & 60.35 & 47.98 & 66.21 \\
 & Self-Planning & 95.12 & 90.24 & 37.00 & 61.14 & 26.93 & 62.09 \\
 & Self-Debugging & \textbf{98.78} & \textbf{96.34} & 61.00 & 82.37 & 74.56 & 82.61 \\
 & INTERVENOR & 95.73 & 92.68 & 63.00 & 79.82 & 70.79 & 80.40 \\
 & \textbf{Ours} & \textbf{98.78} & \textbf{96.34} & \textbf{78.00 ($\uparrow$\,23.81\%)} & \textbf{88.33 ($\uparrow$\,7.24\%)} & \textbf{83.77 ($\uparrow$\,12.35\%)} & \textbf{89.04 ($\uparrow$\,7.78\%)} \\ \midrule
\multirow{6}{*}{\begin{tabular}[c]{@{}l@{}}Qwen-Plus\\-2025-01-25\end{tabular}} & Direct & 90.85 & 86.59 & 31.00 & 50.09 & 41.49 & 60.00 \\
 & CoT & 93.29 & 87.19 & 33.00 & 48.07 & 43.50 & 61.01 \\
 & Self-Planning & 90.85 & 84.75 & 37.00 & 37.36 & 41.75 & 58.34 \\
 & Self-Debugging & \textbf{96.34} & \textbf{93.90} & 49.00 & 70.96 & 63.77 & 74.80 \\
 & INTERVENOR & 95.12 & 91.46 & 48.00 & 68.60 & 61.75 & 72.99 \\
 & \textbf{Ours} & \textbf{96.34} & \textbf{93.90} & \textbf{63.00 ($\uparrow$\,28.57\%)} & \textbf{71.93 ($\uparrow$\,1.37\%)} & \textbf{68.60 ($\uparrow$\,7.57\%)} & \textbf{78.75 ($\uparrow$\,5.28\%)} \\ \bottomrule
\end{tabular}%
}
\end{table*}

\subsection{Performance Evaluation (RQ1)} 
\label{ssec:performance_evaluation}
To address RQ1, we conduct a comprehensive evaluation of TraceCoder against several baseline methods across three foundation models and four benchmark datasets, with results summarized in Table~\ref{tab:main_results_split}. TraceCoder consistently achieves the highest Pass@1 accuracy, outperforming both non-iterative methods (e.g., \texttt{Direct} and \texttt{CoT}) and iterative ones (e.g., \texttt{Self-Debugging} and \texttt{INTERVENOR}). While latter incorporate feedback loops, they are limited by single-path repair strategies and lack mechanisms for learning from prior failures, often resulting in convergence to suboptimal solutions.

TraceCoder overcomes these limitations through multi-agent collaboration and runtime trace-driven analysis, enabling more precise and adaptive repairs. This advantage is especially evident on structurally complex benchmarks such as ClassEval and BigCodeBench. For example, using Gemini-2.5-Flash-0417 on ClassEval, TraceCoder achieves a Pass@1 score of 82.00\%, surpassing the second-best baseline (61.00\%) by a relative margin of 34.43\%. A similar trend is observed on BigCodeBench. When averaged across all benchmarks, TraceCoder reaches 90.72\% with Gemini, outperforming the strongest baseline (81.05\%) by 11.93\%, demonstrating robust performance across diverse programming tasks.

These results provide strong empirical evidence for the effectiveness of TraceCoder’s framework design. By combining runtime instrumentation and historical lesson learning, TraceCoder enables the Analysis Agent to identify root causes rather than surface-level symptoms, leading to more accurate and targeted repairs, particularly for complex bugs that are difficult to resolve.

\setlength{\fboxsep}{8pt}
\setlength{\fboxrule}{0.6pt}%
\noindent\fcolorbox{gray!75!black}{gray!5!white}{%
  \begin{minipage}{\dimexpr\linewidth-2\fboxsep-2\fboxrule}%
    \textbf{Answer to RQ1:} TraceCoder consistently outperforms baseline methods across all settings. Its advantage is particularly notable on complex benchmarks such as ClassEval and BigCodeBench, achieving a relative improvement of up to 34.43\% over the strongest baselines.
  \end{minipage}%
}

\subsection{Impact of Framework Parameters (RQ2)}
\label{ssec:parameter_impact}
\begin{figure}
  \centering
  \includegraphics[width=\columnwidth]{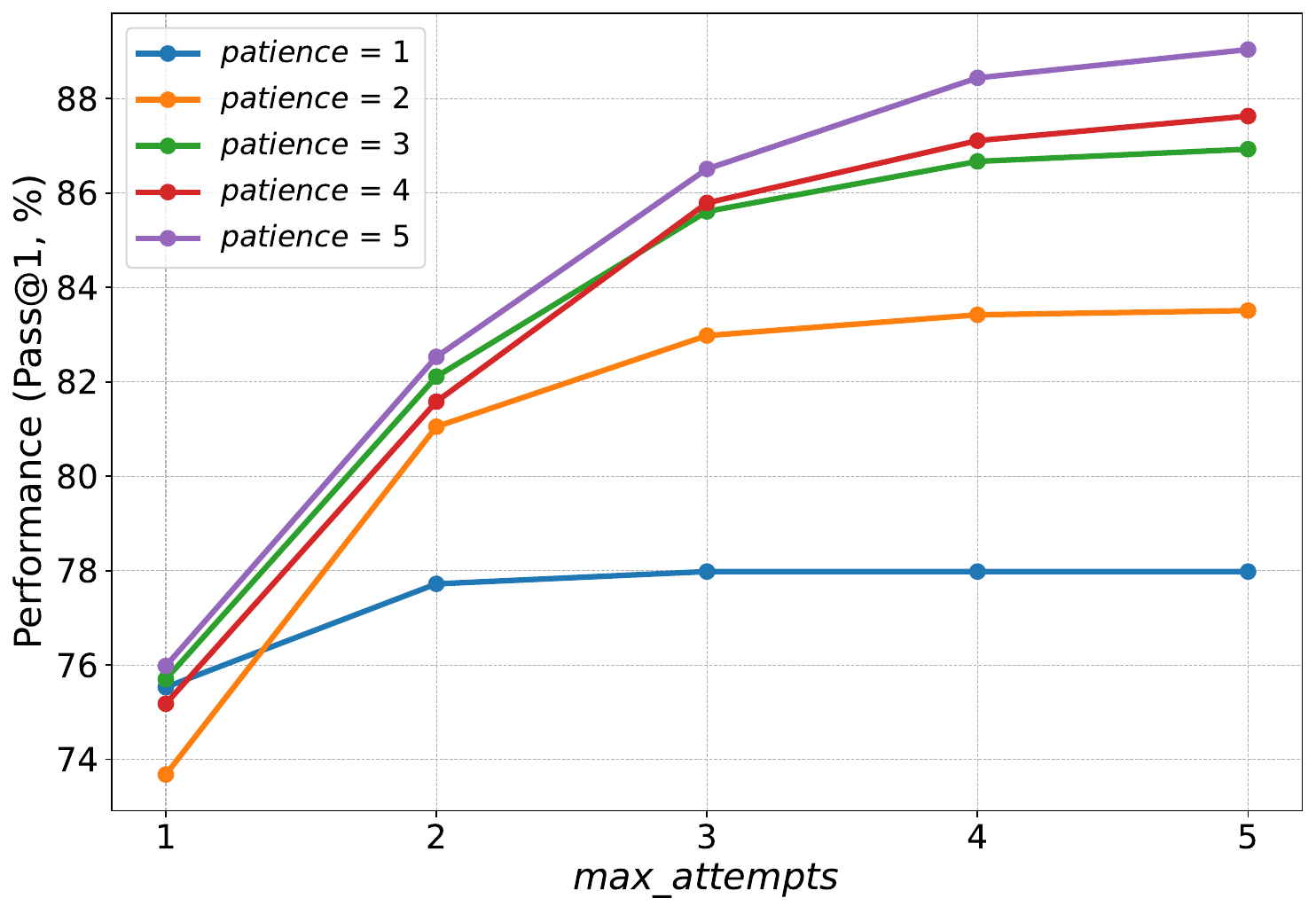}
  \Description[Hyperparameter sensitivity curve]{
    Sensitivity plot showing how varying max_attempts and patience affects the
    Pass@1 performance of Gemini-2.5-Flash-0417 on BigCodeBench‑Complete dataset.
  }
  \caption{Sensitivity analysis of \textit{max\_attempts} and \textit{patience} on Pass@1 performance. Results are reported on BigCodeBench-Complete using Gemini-2.5-Flash-0417.}
  \label{fig:param_sensitivity}
\end{figure}

We analyze RQ2 by examining the impact of two key hyperparameters: \textit{max\_attempts}, which defines the upper limit of repair iterations, and \textit{patience}, which controls the early stopping threshold. Sensitivity analysis is conducted on BigCodeBench-Complete using Gemini-2.5-Flash-0417, with results shown in Figure~\ref{fig:param_sensitivity}.

\paragraph{Effect of \textit{max\_attempts}} 
We observe a consistent increase in Pass@1 accuracy as \textit{max\_attempts} increases, confirming the benefit of iterative refinement in TraceCoder’s design. Allowing more repair cycles offers additional opportunities for diagnosis and repair.

\paragraph{Effect of \textit{patience}} 
Similarly, increasing \textit{patience} yields steady performance gains. A higher tolerance enables the framework to perform multi-step repairs and better utilize the LLM’s stochastic exploration, especially when early attempts fail. Peak performance is reached at the highest tested \textit{patience} level, suggesting that greater fault tolerance enhances effectiveness on complex tasks. However, this improvement comes at the cost of increased computational overhead, indicating a trade-off between accuracy and efficiency.

\setlength{\fboxsep}{8pt}
\setlength{\fboxrule}{0.6pt}%
\noindent\fcolorbox{gray!75!black}{gray!5!white}{%
  \begin{minipage}{\dimexpr\linewidth-2\fboxsep-2\fboxrule}%
    \textbf{Answer to RQ2:} TraceCoder's performance is highly sensitive to both \textit{max\_attempts} and \textit{patience}. Increasing either improves accuracy by enabling the framework to escape local optima. For complex tasks, greater \textit{patience} enables deeper exploration, leading to more successful repairs.
  \end{minipage}%
}

\subsection{Ablation Study (RQ3)}
\label{ssec:ablation_study}
To evaluate the individual contributions of TraceCoder’s core components, we conduct an ablation study on the BigCodeBench-Complete dataset. The results are detailed in Table~\ref{tab:ablation_study}. We systematically disable key parts of the framework to assess their impact.
\begin{itemize}
    \item \textbf{w/o Instrumentation}. Removes the Instrumentation Agent. The Analysis Agent must operate without runtime traces.
    \item \textbf{w/o Instrumentation \& Analysis}. Excludes both Instrumentation and Analysis Agents, leaving the Repair Agent to act solely on raw test failure feedback.
    \item \textbf{w/o Iterative Repair}. Disables the entire agent-based repair framework, reverting the system to single-pass code generation without iterative reasoning.
    \item \textbf{w/o HLLM}. Deactivates HLLM, which eliminates the framework’s ability to incorporate lessons from prior failures.
    \item \textbf{w/o RM}. Removes the RM, requiring the framework to proceed from the latest attempt regardless of its performance.
    \item \textbf{w/o HLLM \& RM}. Disables both HLLM and RM to evaluate their joint contribution to overall performance.
\end{itemize}

\begin{table}
\centering
\caption{Ablation study of TraceCoder's components on the BigCodeBench-Complete dataset. The table shows the Pass@1 performance (\%) after removing specific components.}
\label{tab:ablation_study}
\begin{tabular*}{0.48\textwidth}{@{\extracolsep{\fill}} lr}
\toprule
\textbf{Configuration} & \textbf{BigCodeBench-Complete} \\
\midrule
TraceCoder & \textbf{89.04} \\
\midrule
w/o Instrumentation & 78.51 \\
w/o Instrumentation \& Analysis & 75.09 \\
w/o Iterative Repair & 53.77 \\
\midrule
w/o HLLM & 86.75 \\
w/o RM & 84.55 \\
w/o HLLM \& RM & 84.43 \\
\bottomrule
\end{tabular*}
\end{table}

The ablation results confirm that the full TraceCoder framework achieves the highest accuracy, with all components contributing positively. The most substantial degradation occurs when the entire iterative repair loop is eliminated (\textbf{w/o Iterative Repair}), resulting in a sharp accuracy drop from 89.04\% to 53.77\%.

Gradual removal of agents highlights their individual significance. Excluding the Instrumentation Agent alone causes a marked decline, underscoring the importance of runtime traces for precise diagnosis. Further excluding the Analysis Agent leads to an additional 4.36\% drop, validating its essential role in formulating structured repair plans even in the absence of trace data.

The supporting mechanisms also play complementary roles. Disabling the RM results in a 5.31\% relative decline, validating its effectiveness in preserving promising states and preventing convergence deterioration. Meanwhile, the HLLM contributes a 2.64\% improvement by avoiding repeated failures and accelerating repair convergence. Removing both RM and HLLM confirms that their impacts are largely independent yet synergistic.

Overall, these findings align with the debugging strategies employed by experienced developers: addressing complex bugs requires not only repeated attempts, but also systematic analysis of runtime behavior (enabled by the Instrumentation and Analysis Agents), retention of prior failures (facilitated by HLLM), and disciplined rollback to stable baselines (enforced by RM).

\setlength{\fboxsep}{8pt}
\setlength{\fboxrule}{0.6pt}%
\noindent\fcolorbox{gray!75!black}{gray!5!white}{%
    \begin{minipage}{\dimexpr\linewidth-2\fboxsep-2\fboxrule}%
        \textbf{Answer to RQ3:} The ablation study confirms that all components of TraceCoder are essential for optimal performance. The foundational iterative repair framework provides the largest improvement, while the RM and HLLM contribute vital optimizations, adding 5.31\% and 2.64\% respectively for stability and learning efficiency.
    \end{minipage}%
}

\subsection{Practical Evaluation: Reliability, Efficiency, Cost, and Failure Analysis (RQ4)}
\label{ssec:practical_evaluation}

To assess TraceCoder’s real-world applicability and understand its underlying behavior, we conduct a multi-faceted operational analysis. Specifically, we examine the reliability of its core instrumentation mechanism, measure computational cost, compare its efficiency against sampling-based strategies, diagnose its dominant failure modes, and illustrate its repair process on a challenging bug.

\subsubsection{Validation of Instrumentation Purity}
\label{sssec:validation_purity}
A core assumption of TraceCoder is that the Instrumentation Agent can insert diagnostic probes without altering the program's original semantics. To empirically validate this assumption, we conduct a targeted study as follows: (1) we collect all initially correct samples generated by the \texttt{Direct} baseline on HumanEval and ClassEval; (2) these correct samples are instrumented using our Instrumentation Agent; and (3) the full test suites are re-executed on the instrumented versions to check whether correctness was preserved.

As shown in Table~\ref{tab:semantic_preservation}, TraceCoder achieves a semantic preservation rate of 99.32\%--100\% across both evaluated LLMs, confirming the reliability of our instrumentation strategy. This provides strong empirical evidence that our prompt-guided instrumentation process is highly reliable and rarely introduces new bugs. To further corroborate these findings, we also employ formal verification using CrossHair's \texttt{diffbehavior} tool\footnote{https://github.com/pschanely/CrossHair}. This analysis confirms that the instrumented code is behaviorally equivalent to the original in over 97\% of cases, providing strong formal evidence for the semantic integrity of our instrumentation process. This validation further suggests that TraceCoder’s prompt-engineering strategy can be generalized to other programming languages and LLMs.

\begin{table}
\centering
\caption{Empirical validation of semantic preservation. The preservation rate measures the percentage of initially correct programs that remain correct after instrumentation.}
\label{tab:semantic_preservation}
\resizebox{\columnwidth}{!}{%
\begin{tabular}{@{}llccc@{}}
\toprule
\textbf{Model} & \textbf{Dataset} & \makecell{Initial Correct \\ Samples} & \makecell{Correct After \\ Instrumentation} & \textbf{Preservation Rate} \\ \midrule
\multirow{2}{*}{Gemini-2.5-Flash} & HumanEval & 161 & 161 & 100.0\% \\
 & ClassEval & 44 & 44 & 100.0\% \\ \midrule
\multirow{2}{*}{Qwen-Plus} & HumanEval & 147 & 146 & 99.32\% \\
 & ClassEval & 30 & 30 & 100.0\% \\ \bottomrule
\end{tabular}
}
\end{table}

\subsubsection{Cost Analysis}
We first evaluate the computational cost of TraceCoder and all baselines on the \textit{Gemini-2.5-Flash-0417}, measuring the average token usage per problem. As shown in Table~\ref{tab:token_consumption_comparison}, non-iterative methods like \texttt{Direct} and \texttt{CoT} consume the fewest tokens, but this comes at the cost of lower performance on complex tasks. Iterative baselines demonstrate improved accuracy but incur substantially higher token usage, establishing a clear trade-off between cost and effectiveness. TraceCoder operates in this high-cost, high-performance regime, leveraging its budget for targeted, intelligent repair rather than unguided attempts.

\begin{table}
\centering
\caption{Token consumption comparison of different methods on Gemini-2.5-Flash-0417. Values indicate the average number of tokens used per problem.}
\label{tab:token_consumption_comparison}
\resizebox{\columnwidth}{!}{
\begin{tabular}{lcccc}
\toprule
\multirow{2}{*}{\textbf{Method}} & \multicolumn{2}{c}{\textbf{HumanEval+}} & \multicolumn{2}{c}{\textbf{ClassEval}} \\
\cmidrule(lr){2-3} \cmidrule(lr){4-5}
 & Input tokens & Output tokens & Input tokens & Output tokens \\
\midrule
Direct & 168.35 & 401.59 & 638.46 & 2,264.56 \\
CoT & 176.70 & 489.14 & 679.80 & 2,249.65 \\
Self-Planning & 584.98 & 544.25 & 1,892.09 & 2,363.0 \\
Self-Debugging & 246.67 & 945.26 & 25,906.64 & \textbf{26,744.15} \\
INTERVENOR & \textbf{921.87} & 919.23 & 22,785.2 & 15,182.29 \\
Ours & 900.45 & \textbf{970.60} & \textbf{29,771.90} & 16,264.34 \\
\bottomrule
\end{tabular}
}
\end{table}

\subsubsection{Efficiency Comparison with Sampling}
To ensure a fair comparison against breadth-based search (sampling), we conduct additional experiments under equal token and attempt budgets. As detailed in Table~\ref{tab:equal_attempts} and Table~\ref{tab:equal_tokens}, when baselines like \texttt{CoT} are allowed to generate multiple samples to match TraceCoder's total attempts or token usage, TraceCoder's Pass@1 still significantly outperforms their Pass@k on complex tasks like ClassEval. This demonstrates that TraceCoder's feedback-guided, depth-first repair strategy is more effective than blind sampling for resolving deep logical flaws, as it uses feedback to intelligently navigate the solution space rather than relying on chance.

\begin{table}
\centering
\caption{Performance Comparison under Equal Attempt Setting (Pass@6 for baselines vs. Pass@1 for TraceCoder).}
\label{tab:equal_attempts}
\resizebox{\columnwidth}{!}{%
\begin{tabular}{@{}llccc@{}}
\toprule
\textbf{Model} & \textbf{Method} & \textbf{HumanEval} & \textbf{HumanEval+} & \textbf{ClassEval} \\ \midrule
\multirow{3}{*}{Gemini-2.5-Flash} & Direct (Pass@6) & 98.78\% & 96.34\% & 54.00\% \\
 & CoT (Pass@6) & 98.78\% & 96.95\% & 55.00\% \\
 & \textbf{Ours (Pass@1)} & \textbf{99.39\%} & \textbf{98.17\%} & \textbf{82.00\%} \\ \midrule
\multirow{3}{*}{Qwen-Plus} & Direct (Pass@6) & 94.51\% & 90.24\% & 39.00\% \\
 & CoT (Pass@6) & 96.34\% & 93.29\% & 41.00\% \\
 & \textbf{Ours (Pass@1)} & \textbf{96.34\%} & \textbf{93.90\%} & \textbf{63.00\%} \\ \bottomrule
\end{tabular}
}
\end{table}

\begin{table}
\centering
\caption{Performance Comparison under Equal Token Budget Setting (Pass@k for baselines vs. Pass@1 for TraceCoder).}
\label{tab:equal_tokens}
\resizebox{\columnwidth}{!}{%
\begin{tabular}{@{}llcc@{}}
\toprule
\textbf{Model} & \textbf{Method} & \textbf{HumanEval+} & \textbf{ClassEval} \\ \midrule
\multirow{3}{*}{Gemini-2.5-Flash} & Direct & 93.90\% (Pass@3) & 58.00\% (Pass@15) \\
 & CoT & 92.68\% (Pass@3) & 58.00\% (Pass@15) \\
 & \textbf{Ours} & \textbf{98.17\% (Pass@1)} & \textbf{82.00\% (Pass@1)} \\ \midrule
\multirow{3}{*}{Qwen-Plus} & Direct & 88.41\% (Pass@3) & 46.00\% (Pass@15) \\
 & CoT & 93.29\% (Pass@3) & 53.00\% (Pass@15) \\
 & \textbf{Ours} & \textbf{93.90\% (Pass@1)} & \textbf{63.00\% (Pass@1)} \\ \bottomrule
\end{tabular}
}
\end{table}

\subsubsection{Error Analysis}
To gain deeper insights into TraceCoder’s remaining failure modes, we analyze the types of errors it encounters on the BigCodeBench dataset. As shown in Table~\ref{tab:error_analysis}, outcomes are categorized into Pass, Runtime Error (RE), Wrong Answer (WA), and Time Limit Exceeded (TLE). The data confirms that \textbf{WA} is the dominant failure mode, reaching up to 8.86\%. This indicates that while our trace-driven framework is effective at resolving explicit runtime errors, the remaining challenge lies in correcting subtle logical defects that produce incorrect outputs without crashing.

\begin{table}[h]
\centering
\caption{Error analysis on BigCodeBench subsets.}
\label{tab:error_analysis}
\resizebox{0.48\textwidth}{!}{
\begin{tabular}{lrr}
\toprule
\textbf{Metric} & \makecell[r]{BigCodeBench-\\Complete} & \makecell[r]{BigCodeBench-\\Instruct} \\
\midrule
Pass & 89.04\% & 85.00\% \\
Runtime Error (RE) & 4.23\% & 5.00\% \\
Wrong Answer (WA) & 6.34\% & 8.86\% \\
Time Limit Exceeded (TLE) & 0.39\% & 1.14\% \\
\bottomrule
\end{tabular}
}
\end{table}

\subsubsection{Case Study: Resolving a Semantic Bug}
Given that WA errors are the main challenge, this case study illustrates how TraceCoder is uniquely equipped to diagnose and fix such semantic issues.

\paragraph{Task}
Write a function \texttt{get\_positives(numbers)} that returns a list of strictly positive numbers from the input list.

\paragraph{Incorrect LLM-Generated Code}
The initial code has a common off-by-one logical error, incorrectly including zero in the output.
\begin{lstlisting}[language=Python]
def get_positives(numbers):
    return [x for x in numbers if x >= 0]  # Bug: Should be > 0
\end{lstlisting}

\paragraph{Test Case and Baseline's Dilemma}
The bug is exposed by the test case \texttt{assert get\_positives([0, 1, -1]) == [1]}. The buggy code returns ``[0, 1]'', causing an \texttt{AssertionError: [0, 1] != [1]}. For a black-box method that only sees this final error message, the root cause is ambiguous. It might try various incorrect fixes, such as changing the list order or modifying the numbers, without understanding that the core issue lies in the filtering logic itself.

\paragraph{TraceCoder's Resolution Process}
To showcase TraceCoder's semantic debugging capability, we illustrate how it resolves a subtle logical error that standard sampling-based methods fail to detect.
\begin{enumerate}
    \item[(a)] \textbf{Smart Instrumentation}: Upon detecting an assertion error, the Instrumentation Agent inserts diagnostic probes into the list comprehension’s filtering logic. These probes are purposefully placed to track the evaluation of each element \textit{without altering the program’s semantics}, providing fine-grained visibility into decision-making behavior.
    
    \item[(b)] \textbf{Execution \& Tracing}: Running the test case with the instrumented code yields a fine-grained runtime trace. 
    \begin{lstlisting}[language=Python]
    DEBUG: Checking num=0. Condition 0 >= 0 is True. Appending.
    DEBUG: Checking num=1. Condition 1 >= 0 is True. Appending.
    DEBUG: Checking num=-1. Condition -1 >= 0 is False. Skipping.
    \end{lstlisting}
    
    \item[(c)] \textbf{Analysis \& Localization}: 
    The Analysis Agent inspects the trace and observes that \texttt{0} satisfies the predicate and is appended to the list. By aligning this observation with the task specification (``strictly positive numbers''), it infers the root cause: the condition \texttt{x >= 0} is semantically incorrect, since \texttt{0} is not a positive value.
    
    \item[(d)] \textbf{Targeted Repair}: 
Based on this precise analysis, the Repair Agent receives an unambiguous plan: the filtering predicate for positive numbers is incorrect. The condition should be \texttt{x > 0} rather than \texttt{x >= 0}. The agent then applies this fix and generates the correct patched code.
\end{enumerate}

This case study demonstrates how TraceCoder leverages internal execution visibility to diagnose and fix semantic bugs that are often opaque to methods relying solely on pass/fail signals.

\setlength{\fboxsep}{8pt}
\setlength{\fboxrule}{0.6pt}%
\noindent\fcolorbox{gray!75!black}{gray!5!white}{%
    \begin{minipage}{\dimexpr\linewidth-2\fboxsep-2\fboxrule}%
        \textbf{Answer to RQ4:} TraceCoder achieves a strong cost–performance trade-off, outperforming baselines even under equal token and attempt budgets. Its guided repair strategy is significantly more efficient than unguided sampling, and our analysis shows that WA remains the principal failure mode. The case study further demonstrates TraceCoder’s ability to resolve these subtle logical errors through fine-grained runtime tracing.
    \end{minipage}%
}

\section{Threats to Validity}
\label{sec:threats_to_validity}

In this section, we assess potential threats to validity to ensure a rigorous and balanced interpretation of the results.

\paragraph{External Validity} A primary threat to external validity lies in the high computational cost of TraceCoder. Its detailed, trace-driven framework, while effective, incurs substantial token usage. Although our experiments confirm a superior cost-performance trade-off, its absolute token consumption could hinder its adoption in resource-constrained environments. Designing more lightweight, token-efficient agents is a crucial direction for future work.

\paragraph{Construct Validity} Our evaluation relies on the provided test suites to measure program correctness. TraceCoder's ability to repair bugs is fundamentally constrained by test coverage; inadequate tests may lead to overfitting, where hidden flaws remain unaddressed. Consequently, passing all tests does not guarantee true program correctness, and complementary validation methods may be required to mitigate this threat.

\paragraph{Internal Validity} 
A potential threat to internal validity is data contamination, where evaluation benchmarks may have been included in the LLMs' pre-training data. We mitigated this risk by using the BigCodeBench dataset, which was released after the knowledge cut-off dates of our selected models. While this substantially reduces the likelihood of direct leakage, indirect contamination cannot be entirely ruled out. Nevertheless, all methods were subject to the same experimental conditions, ensuring a fair comparison.

\section{Conclusion}
\label{sec:conclusion}
This paper introduces \textit{TraceCoder}, a trace-driven multi-agent framework that emulates expert debugging behavior to automatically repair LLM-generated code. Through the integration of runtime instrumentation, coordinated agent collaboration, and iterative refinement, TraceCoder enables precise fault localization and targeted correction. Its HLLM prevents redundant failures by reusing past insights, while the RM stabilizes progress across iterations. Extensive evaluations confirm substantial improvements over state-of-the-art baselines, particularly on complex programming tasks.

Future work will focus on improving token efficiency and extending TraceCoder's capabilities. A key direction is scaling the framework to repository-level debugging, which raises challenges such as managing large contexts, mitigating instrumentation explosion, and handling cross-file dependencies. To address these, we plan to explore solutions such as static analysis for coarse-grained localization and retrieval-augmented generation to construct minimal execution contexts for repair. Furthermore, inspired by the trade-off between depth- and breadth-first exploration, we will investigate hybrid strategies that combine initial sampling with parallel TraceCoder instances to efficiently explore a broader solution space. Another promising direction is enhancing the HLLM with structured knowledge representations, enabling agents to generalize repair strategies across different tasks and domains.

\begin{acks}
This work was supported by the Science and Technology Research Program of the Chongqing Municipal Education Commission (Grant No. KJQN202500631), the Humanities and Social Sciences Fund of the Ministry of Education (Grant No. 20YJCZH047), the National Research Foundation Singapore and DSO National Laboratories under the AI Singapore Programme (AISG Award No. AISG4-GC-2023-008-1B), the Fundamental Research Funds for the Central Universities (Grant No. KG16426001), the National Natural Science Foundation of China (Grant No. 62402400), and the National Key Research and Development Program of China (Grant No. 2024YFF0908000).
\end{acks}

\bibliographystyle{ACM-Reference-Format}
\bibliography{ref}

\end{document}